\documentclass[twocolumn,floatfix, groupedaddress,superscriptaddress,altaffilletter,runinaddress]{revtex4}
\usepackage{amsmath}
\usepackage{amssymb}
\usepackage{graphicx}
\usepackage{dcolumn}
\usepackage{graphicx}
\usepackage{dcolumn}
\usepackage{bm}

\def\be{\begin{equation}}
\def\ee{\end{equation}}
\def\ba{\begin{eqnarray}}
\def\ea{\end{eqnarray}}

\begin{document}

\title{Viscous electron flow in mesoscopic two-dimensional electron gas.}

\author{G.~M.~Gusev}
\email{gusev@if.usp.br}
\affiliation{Instituto de F\'{\i}sica da Universidade de S\~ao
Paulo, 135960-170, S\~ao Paulo, SP, Brazil}

\author{A.~D.~Levin}
\affiliation{Instituto de F\'{\i}sica da Universidade de S\~ao
Paulo, 135960-170, S\~ao Paulo, SP, Brazil}

\author{E.~V.~Levinson}
\affiliation{Instituto de F\'{\i}sica da Universidade de S\~ao
Paulo, 135960-170, S\~ao Paulo, SP, Brazil}

\author{A.~K.~Bakarov}
\affiliation{Institute of Semiconductor Physics, Novosibirsk 630090,
Russia}
\affiliation{ Novosibirsk State University, Novosibirsk 630090, Russia}

\date{\today}
\begin{abstract}
We report  electrical and magneto transport measurements in  mesoscopic size, two-dimensional (2D)
 electron gas in a GaAs quantum well. Remarkably, we find that the probe configuration and sample geometry strongly affects
 the temperature evolution of local resistance. We attribute all transport properties  to the  presence of hydrodynamic effects.
Experimental results confirm the theoretically predicted significance of  viscous flow in mesoscopic devices.

 \pacs{73.43.Fj, 73.23.-b, 85.75.-d}

\end{abstract}

\maketitle

In the last two decades, there has been considerable progress in the understanding of electron transport in micro and nanometer scaled systems.
Successful fabrication of ballistic field-effect transistors requires a fundamental understanding of the mechanism of charge carrier transport.
 The commonly accepted mechanism for the transport properties is described semiclassically or by the Landauer-Buttiker formalism.
 Note, however, that these models are based on the assumption that the rate of momentum conserving scattering exceeds that of momentum relaxation
 scattering. It is important to look at different principles for a theory of transport. There has been increasing interest in the fabrication of devices with new types
 of functionality whose operation is determined by new principles. A remarkable possibility is the hydrodynamic regime of a Fermi liquid of electrons in a two-dimensional
 system, when the mean free path for electron-electron collisions $l_{ee}$ is smaller
than the mean free path with static defects and phonons $l$, and  transport resembles a viscous electron fluid [1-8].
The viscosity contribution to the transport can be specially enhanced in a pipe-low set up, where the mean free path $l_{ee}$ is much less than
the sample width $W$, while $l >> W$. In such a hydrodynamic regime, the theory makes a number of dramatic predictions, for example, the resistivity is  inversely proportional to
the square of the temperature, $\rho \sim T^{-2}$, so-called “Gurzhi effect”,  and  the square of the sample width
$\rho \sim W^{-2}$ [1,2]. This effect has not been experimentally observed until now, even where other signatures for hydrodynamics have been demonstrated.
Conventional liquid Fermi theory predicts $\rho \sim T^{2}$, since quasiparticles near the Fermi surface
scatter at a rate $T^{2}$.

In experiments, the viscous 2D electron transport has been examined in electrostatically defined GaAs wires using current heating technique [9,10].
 Recently large negative magnetoresistance has been observed in high mobility 2D gas in GaAs macroscopic samples [11,12].
 However, a significant portion of the attention in hydrodynamic effects has been dedicated to graphene for its very weak scattering against acoustic phonons,
 which allows for the realization of hydrodynamic flow at high temperatures. Indeed several theoretical predictions have been confirmed in high quality, encapsulated,
 single layer graphene: negative vicinity [13] resistances have been observed and successfully explained by vorticity generated
 in viscous flows [14-17]. Note, that such a dramatic experimental appearance of electron viscosity in nonlocal transport has not been accompanied by effects
 in longitudinal resistance and magnetotransport.

 A series of updated theoretical approaches has been published recently [18-21], providing additional possibilities to determine the  viscosity from local and magnetotransport measurements,
 which require experimental verifications.

\begin{figure}[ht]
\includegraphics[width=8cm]{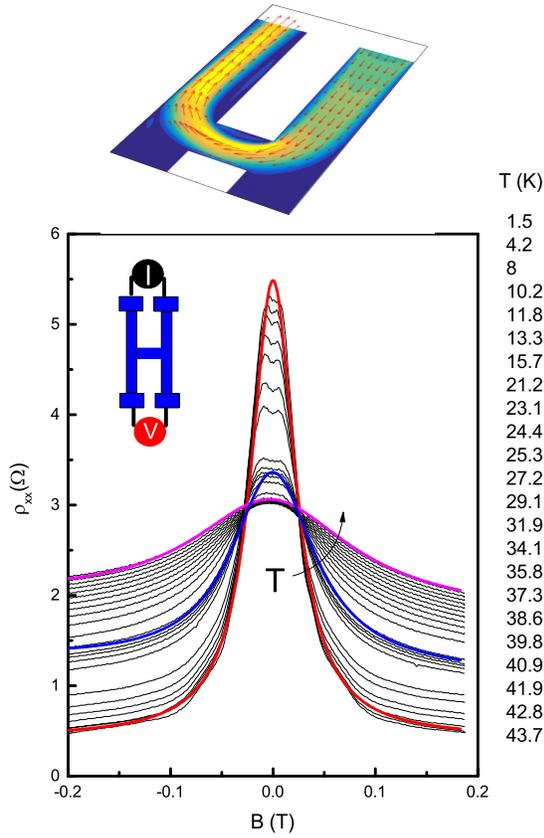}
\caption{(Color online) Top- a sketch of the velocity flow profile for viscous flow in the experimental set up used in this study.
Temperature dependent magnetoresistance of a GaAs quantum well in an H-bar sample. Thick curves
are examples illustrating magnetoresistance calculated from Eqs. 1,2 in main text for  different temperatures: 1.5 K (red), 27,2 K(blue) and 43,7 K( magenta).
The schematics show how the current source and the voltmeter are
connected for the measurements.}
\end{figure}

In the present paper, we have gathered  all the requirements for observation of the hydrodynamic effect in a 2D electron system and present experimental results
accompanied with quantitative analysis. For this purpose, we have chosen GaAs mesoscopic samples  with high mobility 2D electron gas. The finding of  previous studies
[9-12] and theoretical approaches [18-21],  illustrate that it has become necessary to revisit electron transport in high quality GaAs systems.
We  employ commonly used longitudinal resistance and magnetoresistance  to characterize electron shear viscosity, electron-electron
scattering time, and reexamine electron transport over a certain temperature range 1.5-40 K.
One particularly striking observation is the change in the sign of the resistance temperature dependence with changing current injection probe configuration.
Moreover, we observe the “Gurzhi effect” in devices with  H-bar geometry.
The electron-electron scattering time and viscosity are extracted from transport measurements and its temperature dependence in a wide region of temperatures.

\begin{figure}[ht]
\includegraphics[width=8cm]{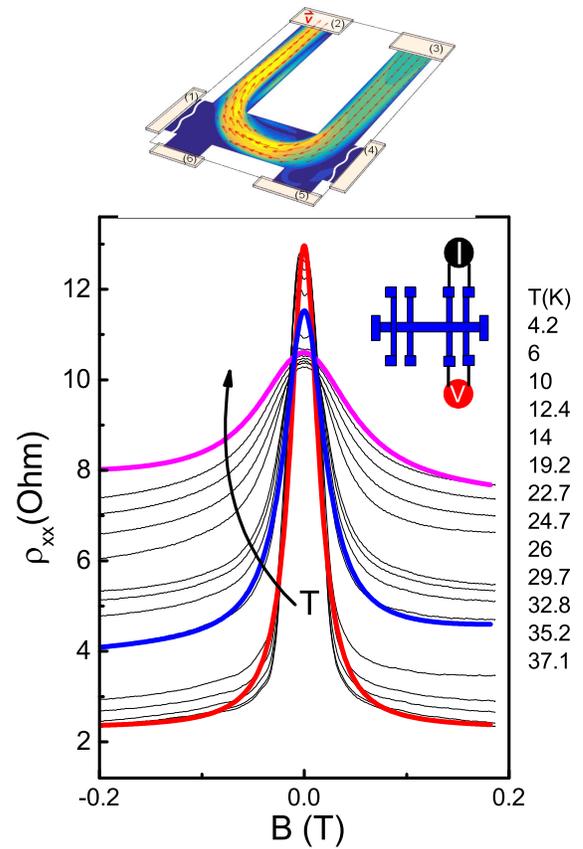}
\caption{(Color online) Top- a sketch of the velocity flow profile for viscous flow in the experimental set up used in this study.
Temperature dependent magnetoresistance of a GaAs quantum well in a Hall bar sample. Thick curves
are examples illustrating magnetoresistance calculated from Eqs. 1,2. for  different temperatures:
4.2 K (red), 19,2 K(blue) and 37,1 K( magenta). The schematics show how the current source and the voltmeter are
connected for the measurements.}
\end{figure}

\begin{figure}[ht]
\includegraphics[width=8cm]{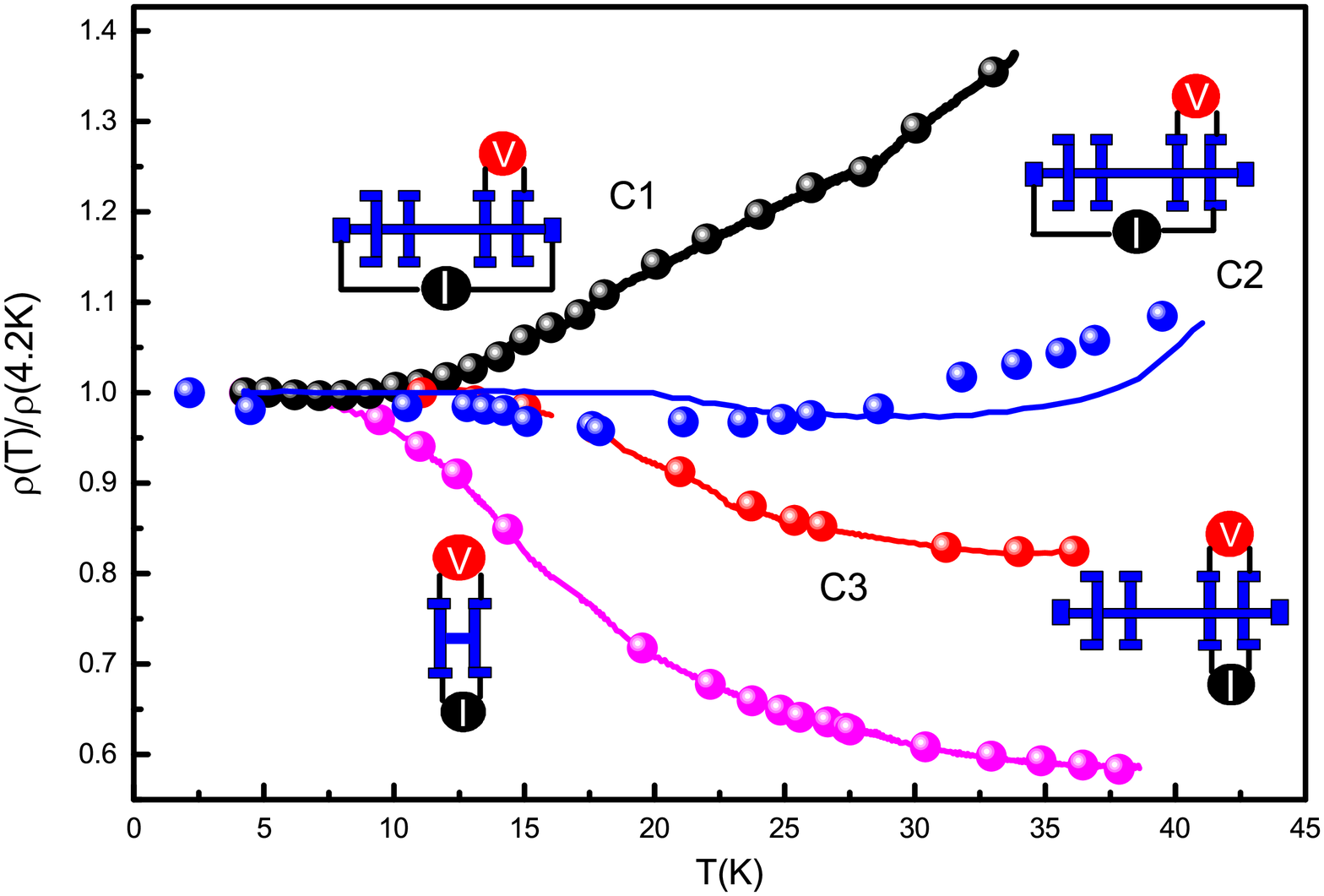}
\caption{(Color online)
Temperature dependent resistivity of a GaAs quantum well in a Hall bar and H-bar for different configurations in zero magnetic field.
Circles show calculations from theoretical formula (1) with numerical parameters described in the main text.
}
\end{figure}

  Our samples are high-quality, GaAs quantum wells with a  width of 14~nm, high
electron density $n_{s}\simeq 9.1\times10^{11}$~cm$^{-2}$, and a mobility of
$\mu~\simeq 2\times 10^{6}$~cm$^{2}$/Vs at $T=1.4$~K . We present experimental results on two different types of mesoscopic size
devices, refereed to as Hall-bar and H-shaped bar, fabricated from the same wafer.
The Hall bar is designed for multi-terminal measurements. The
sample consists of three, $5 \mu m$ wide consecutive segments of
different length ($10, 20 , 10 \mu m$), and 8 voltage probes.
The four terminal, H -shaped bar consist of a $4\times10 \mu m^{2}$
central channel between $5 \mu m$ wide legs.
The measurements were carried out in a
VTI cryostat, using a conventional
lock-in technique to measure the longitudinal $R_{xx}$ resistance with an
ac current of $0.1 - 1 \mu A$ through the sample, which is
sufficiently low to avoid overheating effects.
Two Hall bars and 4 H-shaped devices from the same wafers have been studied. We also compare
our results with transport properties of 2D electrons in a macroscopic sample.

Fig. 1 shows the longitudinal magnetoresistivity $\rho_{xx}$ measured in local configuration for a H-bar sample as a function of magnetic field and temperature.
One can see two characteristic features: a giant negative magnetoresistance $(\sim 400-1000 \%)$ and a pronounced temperature dependence of the zero field resistance.
Surprisingly, the resistance decrease with temperature almost follows $\rho \sim T^{-2}$ dependence, as in the Gurzhi effect.
Fig. 2 shows the longitudinal magnetoresistivity $\rho_{xx}$ measured in local configuration for a Hall-bar sample as a function of magnetic field and temperature.
Note, that we use a set up, where  the current is injected through the system  at a lateral contact (referred as C3 configuration), which resembles current flow in a H-bar
sample. The magnetoresistance feature is qualitatively similar, although the decrease is not so rapid as in the H-bar.
We also check the conventional set up, where current is injected through probe 1 to 4, and the voltage is measured between probes 2  and 3 (referred as C1 configuration)
Strikingly, while in the viscous regime it is expected that electro-electron scattering time $\tau_{ee}$
behaves as $\propto T^{-2}$ in both set ups, resistance increases with T in the conventional measurement set up C1 and decreases with T in
the set up where the current injection probes are positioned against the voltage probes C3. The results for the different schematic set ups
in zero magnetic field are shown in Fig 3. One can see that the temperature coefficient of resistance is strongly affected by probe
configuration.

In mesoscopic samples, two transport regimes can be identified:  ballistic and hydrodynamics.
 In order to distinguish the ballistic and hydrodynamic regimes more in depth
analysis of the problem should be done. Significant temperature dependence of the value and shape of magnetoresistance and dependence on the probe configurations
is inconsistent with dominant ballistic contribution.
\begin{figure}[ht]
\includegraphics[width=8cm]{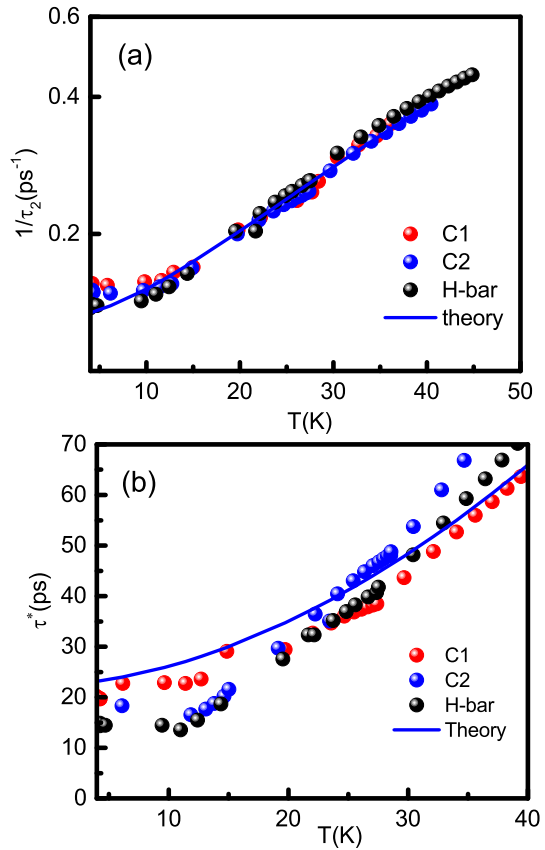}
\caption{(Color online)
(a) The relaxation time $\tau_{2}$ as a function of the temperature obtained by fitting
the theory with experimental results. The solid line is theory.
(b) The relaxation time $\tau^{*}$ as a function of the temperature obtained by fitting
the theory with experimental results. The solid line is theory with parameters presented in the main text.}
\end{figure}

We compare our results with previously published models [18-20]. A more advanced model, however, restricted by
a zero magnetic field, consider both local and nonlocal transport in graphene [17]. The model is generic and can be applied to other material with a parabolic spectrum
such as GaAs quantum wells. The resulting conductivity of 2D gas in constrained geometry is given by
\begin{equation}
\sigma=\sigma_{0}(1-{\cal F}),~~{\cal F}=2\frac{D}{W\xi}\sinh \left(\frac{W}{2D} \right),
%1
\end{equation}
where $\sigma_{0}=e^{2}n\tau /m=1/\rho_{0}$ is the Drude conductivity, $\tau$ is momentum relaxation time due to interaction with phonons and
static defects, $D=\sqrt{\eta \tau_{ee}}$, $\xi=l_{s}\sinh(W/2D)+D\cosh(W/2D)$ is characteristic length
which depends on the boundary slip length $l_{s}$.  The boundary no-slip conditions correspond to the ideal hydrodynamic case of diffusive boundaries with $l_{s}=0$,
while the opposite limit (free surface boundary conditions) corresponds to the ideal ballistic case with $l_{s}=\infty$. Asymptotic limit (ideal hydrodynamic approach)
$l_{s}=0$ has been considered in [18,19] and extended to nonzero magnetic field.  In this case, the conductivity (1) can be substituted by a simple interpolation formula
\begin{equation}
\rho=\rho_{0}\frac{1}{1-2\frac{D}{W}\tanh(\frac{W}{2D})}\approx \rho_{0}\left(1+\frac{\tau}{\tau^{*}}\right),
%2
\end{equation}
where the effective relaxation time is given by [18-20]:
\begin{equation}
\tau^{*}=\frac{W(W+6l_{s})}{12\eta}
%3
\end{equation}
\begin{equation}
\eta=\frac{1}{4}v_{F}^{2}\tau_{2}.
%4
\end{equation}

\begin{equation}
\frac{1}{\tau_{2}(T)}=A_{ee}^{FL}\frac{T^{2}}{[ln(E_{F}/T)]^{2}}+\frac{1}{\tau_{2,0}}
%5
\end{equation}
where the coefficient $A_{ee}^{FL}$  be can expressed via the Landau interaction parameters ($A_{ee}^{FL}=1.5\times10^{10} s^{-1}K^{-2}$), and $\tau_{2,0}$
is the scattering time from disorder.

 Therefore, viscosity leads to incorporation of an extra relaxation mechanism, which contains the
contribution from the electron-electron scattering time $\tau _{2,ee}(T)$ and temperature independent electron scattering from disorder $\tau _{2,0}$  [18,19].
In other words, the small ratio between relaxation of the second moment of electron distribution function and first moment $\tau^{*}/\tau=l_{2,ee}/l<<1$ corresponds
to the dominant viscous contribution to resistivity.
Such separation of the conductivity in two independent channels allows the introduction of the magnetic field dependent viscosity tensor and the derivation of
magnetoresisivity  [18,19]:
\begin{equation}
 \rho_{xx}= \rho_{0}\left(1+\frac{\tau}{\tau^{*}}\frac{1}{1+(2\omega_{c}\tau_{2})^{2}}\right).
%6
 \end{equation}

We fit the magnetoresistance curves in Figs 1 and 2 and resistance in zero magnetic field, shown in fig.3, with the following fitting parameters :
$\tau_{2,0}=0.8\times10^{-11}$ s, $\tau_{0}=10^{-9} s$, $A_{ee}^{FL}=0.9\times10^{9} s^{-1}K^{-2}$.
We also find that in both  microscopic and macroscopic samples $\frac{1}{\tau(T)}=A_{ph}T+\frac{1}{\tau_{0}}$
Assuming that the viscous effect is small in macroscopic samples,
we can reduce the number of independent parameters by measuring $\rho_{0}(T)\sim 1/\tau(T)$  and extract $A_{ph}$ independently. We find $A_{ph}=10^{9}s^{-1}K^{-1}$.

Fig. 4a shows the dependencies of $\tau_{2}(T)$  extracted from comparison with the theory.
Indeed the electron-electron scattering time follows expected behaviour described by equation 5.
The effective relaxation time $\tau^{*}$ is proportional to the second moment relaxation rate $\frac{1}{\tau_{2}}$ (not a time) and
can be also compared with the theory, as we can see from eqs.3 and 4. Note, however, that $\tau^{*}$ contains additional parameter -boundary slip length,
which depends on the viscous flow conditions. We are able to reproduce the evolution of characteristic time with temperature, assuming that $l_{s}$ depends on probe configuration.
We find the value of $l_{s}$ for corresponding set ups and sample geometries: $3.2\mu m$ (C1), $2,4\mu m$ (C2), $1,1\mu m$ (H-bar).
Although it could have been expected that all dependencies merge in a single curve, the curves show a tendency to collapse into one.
The remaining misfitting may be related to temperature dependence of $l_{s}$.
Therefore, the different sign of the temperature coefficient for different set ups is explained by the viscous flow conditions because of the decreasing of $\tau^{*}$ or $l_{s}$.
 It is worth noting that, the dependence of the boundary slip length on the probe configuration and geometry still requires further investigation. We modeled the Poiseuille flow for two dimensional situations depicted in Figs 1 and 2 (top). We find that the velocity profile is strongly depends on the geometry and liquid
  flow injections. Calculation of potential distribution in a viscous charged liquid is a very challenging theoretical task and is out of the scope of the present experimental work.
  Note, however, that more advanced consideration predicts that diffusive scattering on the rough edge and inhomogeneity of the velocity field due to geometry may result in a similar effect [18]. In this case
$\tau^{*}\sim d^{2}/\eta$, where $d$ is the characteristic period of static defects or velocity inhomogeneity [18].

In conclusion, we have measured the evolution of several  magnetotransport characteristics in high quality GaAs quantum wells with temperature. In order to fulfill
   requirements for a hydrodynamic regime, we use mesoscopic samples, where very recently numerous different predictions have been made [18-21].
These results open up possibilities to control the current flow in microstructures by variation of the viscosity and manipulation of the fluids at a
 micro and nanoscale, developing new  microtechnologies.

We thank Z.D.Kvon for helpful
discussions. The financial support of this work by FAPESP, CNPq
(Brazilian agencies) is acknowledged.

\end{document}